\documentclass[prb,showpacs,byrevtex,twocolumn]{revtex4}
\usepackage{epsf}
\usepackage{epsfig}
\usepackage{amssymb}

\newcommand{\be}{\begin{equation}}
\newcommand{\ee}{\end{equation}}
\newcommand{\bea}{\begin{eqnarray}}
\newcommand{\eea}{\end{eqnarray}}

\begin{document}
 \title{Underscreened Kondo impurities in a Luttinger liquid}
\author{
P. Durganandini$^{1}$ and  Pascal Simon$^{2}$
}
\affiliation{$^1$Department of Physics,
              University of Pune,
              Pune 411 007, India.
\\
$^2$Laboratoire de Physique et Mod\'elisation des Milieux Condens\'es, CNRS et Universit\'e
Joseph Fourier, 38042 Grenoble, France
}
\date{\today}

\begin{abstract} 
We study the problem of underscreened Kondo physics in an interacting electronic system modeled by a Luttiger Liquid (LL).
 We find that the leading temperature dependence of thermodynamical quantities 
like the specific heat, spin susceptibility are Fermi Liquid like in nature. 
However, anomalous  power law exponents are seen in the subleading terms. We also discuss possible 
realizations through single and double quantum dot configurations coupled to LL
leads and its consequences for electronic transport. The leading low 
temperature transport behavior is seen to exhibit in general, non Fermi 
liquid LL behavior unlike the thermodynamical quantities. 
 
\end{abstract}

\maketitle
      
\section{Introduction}
   There has been a resurgence of interest in the study of underscreened 
Kondo models in recent years due to their possible role in the observed 
breakdown of Fermi liquid behavior in the neighborhood of a quantum critical 
point in many heavy fermion  materials \cite{si1, coleman1, si2, pepin, mehta}
 as well as the possibility that such underscreened models may be realized for 
quantum dot configurations.\cite{coleman2, pustilnik, hofstetter}
Although the thermodynamics of these models are well known, the dynamical 
properties have been studied only recently.\cite{coleman1} In particular,
it has been emphasized that at zero temperature, the presence of free spins in 
the underscreened models gives rise to singular scattering leading to what 
has been termed as 'singular' Fermi liquid behavior.\cite{mehta} Electronic 
transport through quantum dots which has parameter regimes with underscreening 
 have also been studied.\cite{coleman2, pustilnik, hofstetter} Correlations 
between the electrons can modify impurity effects quite dramatically. 
An example is provided by interacting electrons in one dimension ($1 D$). Such systems have the property that any arbitrary Coulomb repulsion between the electrons generically drives 
the system away from Fermi Liquid (FL) to a Tomonaga-Luttinger liquid (LL)\cite{LL1, LL2, LL3} behavior. In the low energy limit, the charge and spin degrees of freedom are separated and described by collective charge and spin density excitations, 
each moving with a characteristic Fermi velocity. As a result, electron 
correlations function show spin charge separation as well as anomalous power 
law dependences. Such one dimensional 
Luttinger liquids can be realized as very narrow quantum wires\cite{wire} or edge states in fractional quantum Hall liquids\cite{FQHE} or single walled carbon nanotubes,\cite{SWNT} etc. The effects of scalar impurities in a LL have 
been well studied and shown to lead to effects like `breaking' or `healing' of 
the $1 D$ chain.\cite{kane} The problem of a spin $1/2$ 
magnetic impurity in a LL has also been largely studied.\cite{toner, furusaki, johan,durga,lehur,eggert,furusaki2,chandra, meden} It has been shown that while the ground state is a singlet 
state just like for the ordinary Kondo problem, the LL properties of the 
conduction electrons show up in the anomalous power law scaling for the Kondo 
temperature as well as the thermodynamics. An interesting question to ask is 
how underscreened Kondo physics manifests itself in a LL. In this paper, we 
study the problem of underscreened Kondo physics in a LL using 
boundary conformal field theory methods to analyze the renormalization group 
flows and to obtain the thermodynamical properties. We find that the leading 
temperature dependence of thermodynamical quantities like the specific heat,
spin susceptibility is FL like in nature. However, the anomalous LL
power law exponents are seen in the subleading terms. We also discuss possible 
realizations through single and double quantum dot configurations coupled to LL
leads and the consequences for electronic transport. The low temperature 
transport behavior is seen to exhibit non Fermi liquid behavior unlike the 
thermodynamical quantities.

 The plan of our paper is as follows. In Sec.~\ref{sec:FP}, 
we analyze the renormalization group flows of the effective low energy model 
using boundary conformal field theory methods and obtain the thermodynamical
properties. In the next section (Sec.~\ref{sec:QD}), we discuss possible
realizations through single and double dot configurations coupled to LL.
We then discuss electronic transport through such systems. We conclude by
summarizing our results.

\section{\label{sec:FP}Fixed point analysis: A boundary conformal field theory approach}
  In the low energy, long wavelength limit, interacting electrons 
moving in a finite size $1 D $ space extending from $-L$ to $L$ can be described by the linearized continuum Hamiltonian with a four Fermi interaction:
\be
H_0 = \int ^{L}_{-L} dx [i v_F \psi _{\sigma} ^{\dag}(x) \partial _x \psi _{\sigma}(x) + U (\psi _{\sigma} ^{\dag}(x) \psi _{\sigma} (x))^2],   
\ee
where $v_F$ is the Fermi velocity and $U$ denotes the strength of the repulsive density-density interaction.
The one dimensional fermion field $\psi _{\sigma} (x)(\sigma =\uparrow,\downarrow )$ can be expanded about the Fermi points $\pm k_F$ in terms of the left moving and right moving fields as 
\be
\psi_{\sigma}= e^{ -i k_{F}x} \psi_{L \, \sigma} (x) +
                                 e^{ ik_{F}x} \psi_{R \,\sigma} (x).
 \ee
The left and right moving fermions may be bosonized as~\cite{affleck}
         \bea
          \psi_{L \uparrow/\downarrow } \sim  \exp -i \sqrt{\frac{\pi}{2}}
   \left({\sqrt{g_c}} {\phi_c} - \frac{1}{{\sqrt{g_c}}}{\tilde{\phi}}_c \pm 
      {\sqrt{g_s}} {\phi_s} \pm \frac{1}{{\sqrt{g_s}}}{\tilde{\phi}}_s\right),  \\      
          \psi_{R \uparrow/\downarrow}   \sim   \exp\,  i \sqrt{\frac{\pi}{2}}
   \left({\sqrt{g_c}} {\phi_c} - \frac{1}{{\sqrt{g_c}}}{\tilde{\phi}}_c \pm 
      {\sqrt{g_s}} {\phi_s} \mp \frac{1}{{\sqrt{g_s}}}{\tilde{\phi}}_s\right).  
\eea
Here $\phi_{\,c,s},\tilde \phi_{\,c,s}$ are linear combinations of the bosons 
$\phi_{L,\uparrow \,\downarrow },\phi_{R,\uparrow \,\downarrow }$ introduced to
represent the fermion fields  $\psi_{L,\uparrow \,\downarrow},\psi_{R,\uparrow \,\downarrow}$:
\bea
\phi_c & \sim & \frac{1}{\surd 2}(\phi_{L,\uparrow}+ \phi_{R,\uparrow}+ \phi_{L,\downarrow}+ \phi_{R,\downarrow}), \\
\tilde \phi_c & \sim & \frac{1}{\surd 2}(\phi_{L,\uparrow} - \phi_{R,\uparrow} + \phi_{L,\downarrow}- \phi_{R,\downarrow}), \\
\phi_s & \sim & \frac{1}{\surd 2}(\phi_{L,\uparrow}+ \phi_{R,\uparrow}- \phi_{L,\downarrow}- \phi_{R,\downarrow}), \\
\tilde \phi_s & \sim & \frac{1}{\surd 2}(\phi_{L,\uparrow} - \phi_{R,\uparrow} - \phi_{L,\downarrow}+ \phi_{R,\downarrow}). 
\eea
In the absence of an external magnetic field, the parameter $g_s=1$. 
The parameter $g_c$ takes the value $1$ for free fermions and has an $U$ 
dependent value less than $1$ for repulsive interaction. The low energy effective bulk Hamiltonian for the 
interacting fermions can be written then in terms of a free  theory of charge and spin bosons with the interactions parametrized by $g_c$  and $g_s$ and moving with Fermi velocities $v_c$ and $v_s$ respectively as
\be
 H_0 = \frac{1}{2} \sum_{\alpha = c,s} v_{\alpha}  \int _{-L} ^{L} dx \,\,\partial_{\mu} \phi_{\alpha}\partial^{\mu} \phi_{\alpha}.
 \ee
Let us now consider the effect of a magnetic impurity of magnitude $S >1/2$
placed at the origin. We can describe the interaction of the
impurity spin $\vec S$  with the conduction electrons at the site $0$ through the 
spin exchange interaction: 
\bea
 H_K &=& J_K \psi ^{\dag} (0) \frac{\vec \sigma}{2} \psi (0)\cdot\vec S \\ \nonumber
&=& J_K[\psi _L^{\dag} (0) \frac{\vec \sigma}{2} \psi _L(0)  + \psi _R^{\dag} (0) \frac{\vec \sigma}{2} \psi _R(0)\\\nonumber
& +& \psi _L^{\dag} (0) \frac{\vec \sigma}{2} \psi _R(0) + \psi _R^{\dag} (0) \frac{\vec \sigma}{2} \psi _L(0)]\cdot\vec S, \label{eq:kondo}
\eea
where the two terms in the second line of Eq. (\ref{eq:kondo}) describe forward scattering and the 
terms in the third line of Eq. (\ref{eq:kondo}) describe backward scattering. 
Finally, $J_K$ is the Kondo coupling.

\subsection{\label{sec:Free fermions}CFT analysis for free fermions}

In the following we briefly recall some results for the corresponding problem 
with non-interacting fermions.\cite{affleck} For free fermions, it is convenient to impose the 
boundary conditions $\psi(-L)= \psi(L)$ and define a parity definite even-odd 
basis:
$\psi _{e(o), L/R} (x) = \psi_{L/R} (x) \pm \psi _{R/L} (-x), x >0$. 
The fermion fields satisfy the boundary conditions $\psi_{e(o),L} (0) = \pm \psi_{e(o),R} (0)$. In this basis, the Hamiltonian can be written as
\bea
H=H_0 +H_K = \int ^{L}_{0} dx i v_F [\psi _{e, L, \sigma} ^{\dag}(x) \partial _x \psi _{e, L, \sigma}(x)
\nonumber \\ 
+ \psi _{o,L, \sigma} ^{\dag}(x) \partial _x \psi _{o,L, \sigma}(x)] 
+ J_K \psi ^{\dag} _{e \,L} (0) \frac{\vec \sigma}{2}\psi_{e \,L} (0)\cdot\vec S.
\eea
Thus, the odd channel electrons decouple from the interaction and the theory 
can be described entirely in terms of the left moving even channel electrons 
on the $1 D$ space $0$ to $L$ with the Kondo interaction at the origin.
The problem reduces therefore to that of the usual single channel Kondo 
problem interacting with a spin $S$ impurity. In the absence of the impurity,
the free fermion theory can be described by the $SU(2)_{c, k=1} \times SU(2)_{s, k=1} WZW $ model with certain specified `gluing' conditions
for the charge and spin degrees of freedom. The Kondo 
interaction is a local interaction involving only the spin degrees of freedom. 
The renormalization group equations tell us that the Kondo interaction is
 marginally relevant for antiferromagnetic (AFM) coupling while it is
 marginally irrelevant for ferromagnetic (FM) coupling. The weak coupling 
fixed point is therefore stable for FM coupling but unstable for AFM  
coupling. For AFM Kondo coupling, the theory flows to the strong coupling (SC)
 fixed point $J_K = \infty$ with the Kondo scale set by  $T_k= D\exp
 -(1/J _K \rho)$ (assuming a constant density of states $\rho$ for the conduction electrons). In the $J _K = \infty$ limit, the ground state can be understood 
in terms of the Nozi{\a`e}res-Blandin\cite{nozieres} picture of quenching 
of part of the impurity spin by the conduction electrons which leads to a $\pi/2$ phase shift for the conduction electrons. The $\pi/2$ phase shift corresponds
to a change in the boundary conditions for the even channel fermions $\psi _{e, L} (0) = - \psi _{e, R}(0)$. Therefore, the strong coupling FP
theory corresponds to that of a decoupled impurity spin of magnitude $s=S-1/2$ 
and a free fermion theory with renormalized boundary conditions.  The renormalization of the boundary conditions in the strong coupling limit leads to a 
modification of the `gluing' conditions for the charge and spin degrees of freedom which correspond here simply to `fusion' with the spin $1/2$ WZW primary 
field in the spin sector.\cite{affleck} Such a renormalization 
of the effects of a local interaction into boundary conditions lies at the 
heart of the boundary critical phenomena. If the boundary 
condition renormalizes to a fixed point (FP), then the effective theory may be 
described by the appropriate boundary conformal field theory (BCFT). The 
operator content of the BCFT can be obtained by imposing modular invariance on 
the theory. The stability as well as the physics around the FP can be 
determined by analyzing all possible perturbations near the FP with the boundary operators.\cite{affleck}

The high temperature or the weak coupling limit physics is governed by the marginally relevant Kondo interaction. Standard perturbative methods can be used to 
obtain the behavior of various physical quantities like the entropy, specific 
heat, spin susceptibility, etc ~\cite{hewson, affleck} which show as expected,
a logarithmic divergence at temperature $T=T_K$ for AFM coupling. 
In the low temperature or  strong coupling limit for AFM coupling, the leading 
perturbation around the strong coupling FP is that of a 
ferromagnetic spin exchange coupling between the leftover spin 
$s = S -1/2$ impurity and the phase shifted conduction electrons via virtual 
nearest neighbor hoppings.\cite{affleck} Since the residual ferromagnetic 
Kondo coupling is marginally irrelevant, the strong coupling fixed point is stable. Leading corrections to the zero temperature entropy can be obtained by a 
perturbative calculation in the marginally irrelevant residual FM coupling.\cite{affleck, hewson} This gives the low temperature entropy as
\bea 
&&S_{imp} (T\ll T_K) = \ln (2s +1) -\frac{\pi ^2}{3}s(s+1) (2\lambda \rho) ^3[1-
 \nonumber \\&& (2 \lambda \rho) \ln (T/T_K)+ 6 (2 \lambda \rho)^2 \ln ^2 (T/T_K) + \dots],\label{eq:sing1}
\eea
where the first term denotes the degeneracy of the residual impurity spin and 
$\lambda$ denotes the strength of the FM coupling between the leftover impurity spin and the conduction electrons. Usual scaling arguments show that $\lambda \rho$ scales as $\lambda \rho \sim \frac{1}{\ln (T/T_K)}$.
The specific heat then has the leading temperature dependence 
\be 
C_{imp} (T\ll T_K)= \pi ^2 s(s+1) \frac{1}{[\ln (T/T_K)]^4} + \dots .\label{eq:sing2}
 \ee
In the presence of a weak magnetic field, the impurity spin susceptibility can 
be computed as 
\be 
\chi_{imp} (T\ll T_K)= \frac{ (g \mu _B)^2 s(s+1)}{3T} \left[1 - \frac{1}{[\ln (T/T_K)]} + \dots \right].\label{eq:sing3}
 \ee 
Thus the marginal exchange coupling between the residual free impurity spin and
the conduction electrons leads to the `singular' Fermi liquid behavior.\cite{mehta}

If a magnetic field $H$ is added, at low temperature $T\ll H\ll T_K)$, the residual impurity spin becomes polarized 
and the ground state degeneracy is lifted. Since there are no impurity spin 
fluctuations, there is no FM coupling between the residual impurity spin and 
the conduction electrons.
 The leading boundary perturbation is now the spin 2 object with dimension $2$ 
just as in the ordinary Kondo problem:\cite{affleck}
 \be
\lambda_2 (\psi_{e,L\uparrow} ^{\dag} \psi_{e, L\downarrow} )^2 ,
\ee
which leads to the usual regular FL behavior for the various physical quantities.

\subsection{\label{sec:LL} CFT analysis for interacting electrons}

 It is not possible in general to
describe the boundary conditions for the interacting electron problem in a
simple way as for the free fermion theory, however, the possible conformally invariant
boundary conditions for the interacting electron theory with a magnetic
impurity (see Eq.(\ref{eq:kondo})) turn out to be particularly simple within the bosonic language - the
only conformally invariant boundary conditions being either the Dirichlet or
Neumann boundary conditions. The bulk theory (in the absence of the Kondo 
interaction) can be identified with both the charge and spin bosons satisfying 
the  Neumann boundary conditions.\cite{wong,durga} The operator content around this fixed point can be
identified, it turns out the  the backscattering component of the electron
spin operator in Eq.(\ref{eq:kondo}) is the lowest dimensional parity invariant
operator which can couple to the impurity spin. This operator
has dimension $(1+g_c)/2$ ~\cite{durga} which is less than $1$ for $g_c <1$. 
Hence this term is relevant for either sign of the Kondo coupling. The weak 
coupling fixed point is therefore unstable for both ferromagnetic and antiferromagnetic perturbations and flows to the strong coupling fixed points $J _K = + \infty$ for AFM coupling and $J _K = -\infty$ for FM coupling. The corresponding Kondo scale is given by $T_K \propto|J_K \rho|^{2/(1-g_c)}$.\cite{toner} At the AFM SC FP,
one can argue using the usual Nozi{\a`e}res and Blandin picture that the impurity 
spin gets locked with the electron at site $0$ forming an effective spin of 
magnitude $s=S - 1/2$ which gets decoupled from the rest of the chain. The 
effective theory therefore becomes that of an open chain with one site removed 
and a decoupled impurity spin of magnitude $s=S -1/2$. At the FM SC FP, the impurity spin is ferromagnetically coupled to the electron at site $0$ to form an  effective spin $S + 1/2$ which in turn couples with the electrons at the sites $-1$ and $+1$ to form an effective spin $s=S-1/2$. The effective theory is that of an open chain with three sites removed and a decoupled impurity spin of magnitude $s=S -1/2$. Thus, in the $L\rightarrow \infty$ limit, both the AFM and FM SC FP are described by an effective theory of two
decoupled semi-infinite LL and a decoupled spin of magnitude $s=S-1/2$. The two decoupled semi-infinite
LL can be described by a BCFT with Dirichlet boundary conditions on the charge and spin bosons.\cite{wong,durga} We now determine the stability of the SCFP by analyzing all possible perturbations around the fixed point. The two decoupled channels can interact with each other and with the remaining spin $S-1/2$ impurity via the boundary operators. From the boundary operator content,\cite{wong} one can see that the lowest dimensional boundary interactions which can occur are:
 
i) the spin exchange coupling between the boundary spin current operator in each decoupled chain and the leftover free impurity spin of size $s = S-1/2$: 
\be
\lambda_1 \left[ \psi_{L, 1} ^{\dag} \frac{\vec \sigma}{2} \psi_{L,1} + \psi_{L,2} ^{\dag} \frac{\vec \sigma}{2} \psi_{L,2} \right]\cdot \vec {s},
\ee   
 with dimension $1$. Since this coupling is generated as in the usual FL case 
by virtual hopping's between the nearest  neighbor site electrons (next nearest neighbor site electrons for FM Kondo coupling) and the decoupled leftover impurity spin, it 
is ferromagnetic in nature and hence marginally irrelevant, 

ii) the hopping of fermions between the two channels via spin flip scattering with the leftover impurity spin: 
\be
\lambda_2  \left[\psi_{L, 1} ^{\dag}\frac{\vec \sigma}{2} \psi_{L, 2} + \psi_{L, 2} ^{\dag}\frac{\vec \sigma}{2} \psi_{L,1}\right]\cdot\vec {s},
\ee
 which has dimension $(1+g_c)/2g_c$ and is irrelevant for $g_c <1$,

iii) the hopping of a fermion between the two channels:
\be
\lambda_3  \left( \psi_{L\uparrow ,1} ^{\dag} \psi_{L\uparrow ,2} + \psi_{L\downarrow ,1} ^{\dag}\psi_{L\downarrow ,2}\right) + \, H.c.,
\ee
with dimension $ (1+g_c)/2g_c$,

iv) the hopping of a charge two spin singlet between the two channels: 
\be
\lambda_4 \left( (\psi_{L \uparrow} ^{\dag} \psi_{L \downarrow} ^{\dag})_1 (\psi_{L\uparrow} \psi_{L\downarrow} )_2 \right) + \, H.c.,
\ee 
with dimension $2/g_c$,
          
v) the hopping of a charge neutral spin $2$ object between the two channels:
\be
\lambda_5 \left( (\psi_{L\uparrow} ^{\dag} \psi_{L\downarrow})_1 (\psi_{L\uparrow} ^{\dag} \psi_{L\downarrow})_2 \right)  + H.c. ,
\ee
 which  has  dimension $2$,

vi) a potential scattering term:
\be
\lambda _6\left( ( \psi_{L\uparrow} ^{\dag} \psi_{L\downarrow} )_1 + (\psi_{L\uparrow }^{\dag} \psi_{L\downarrow} )_2\right),
\ee
  which has dimension $1$, 
 
vii) and a spin two object:
\be
\lambda _7 \left[ ((\psi_{L\uparrow} ^{\dag} \psi_{L\downarrow} )_1 )^2  + ((\psi_{L\uparrow} ^{\dag} \psi_{L\downarrow} )_2 )^2 \right],
\ee
which has dimension $2$.

  The potential scattering term is an exactly marginal operator and can only 
lead to a shift of the ground state energy. Since all these operators are 
irrelevant for  $g_c<1$, the fixed  point  is stable to these perturbations.

  We next discuss the physics around the weak and strong coupling FP.
The Kondo backscattering term governs the physics near the weak coupling FP.
For $T>>T_K$, the leading temperature dependence of the entropy, specific heat 
and the impurity spin susceptibility can be obtained as follows:
\begin{widetext}
\be
S_{imp} (T\gg T_K)= \ln (2S +1) + A S(S+1) (T_K/T) ^{(1-g_c)} + \dots,
 \ee
 \be
C_{imp}(T\gg T_K) = A (g_c-1)S(S+1)(T_K/T)^{(1-g_c)} + \dots,
\ee
\be
\chi _{imp} (T\gg T_K) = \frac{ (g \mu _B)^2 S(S+1)}{3T} [1 - B (T_K/T) ^{(1-g_c)}+ \dots],
\ee
where $A$ and $B$ are non-universal dimensionless constants depending on $g_c$ 
and the electron density of states and the dots denote subleading terms.

  The lowest dimension boundary perturbation near the strong coupling FP is 
the marginally irrelevant exchange coupling between the boundary spin current 
operator in each channel and the residual impurity spin. The next lowest 
dimensional boundary perturbation is the electron hopping term between the 
two channels via spin flip scattering with the residual impurity spin. The 
leading corrections to the low temperature entropy can be expressed in terms 
of the irrelevant coupling parameters $\lambda_i$ with $i=1,\cdots 7$ (redefined in terms of 
dimensionless quantities) as follows:
\be
S_{imp} (T\ll T_K)= \ln (2s +1) -\frac{\pi ^2}{3}s(s+1)[\lambda _1 ^3 + c_2 \lambda _2 ^2 + ...] + c_3 \lambda _3 ^2 + \dots,
\ee
The first term in the above equation denotes the degeneracy of the left-over 
impurity spin, the next two terms are due to the two lowest dimension boundary 
operators interacting via the residual impurity spin and the dots inside the 
bracket  indicate subleading terms due to interactions with the residual 
impurity. The next term indicates the boundary contribution from electron 
tunneling between the two channels without interaction with the residual 
impurity spin and the final dots indicate higher order contributions from 
residual spin impurity independent boundary operators. It is easy to see from 
the scaling dimensions of the boundary operators that $\lambda _1$ scales as 
$\lambda _1 \sim \frac{1}{\ln (T/T_K)}$ while 
$\lambda _2$ and $\lambda _3$ scales as $(T/T_K)^{(1-g_c)/2g_c}$. The temperature behavior of the specific heat can be obtained as follows:
\be
C_{imp} (T\ll T_K) =  \pi ^2 s(s+1)[\frac{1}{[\ln(T/T_K)]^4} + c_2  (T/T_K)^{(1-g_c)/g_c}+ ...] + c_3 (T/T_K)^{(1-g_c)/g_c}+ \dots.
 \ee
Similarly, the zero field impurity spin susceptibility is given as follows:
\be
\chi_{imp} (T\ll T_K) = \frac{ (g \mu _B)^2 s(s+1)}{3T} [1 - \frac{1}{\ln (T/T_K)} + 
c_2 (T/T_K)^{(1-g_c)/g_c}  + ..] + c_0 +\dots,
\ee
\end{widetext}
$c_0, c_2, c_3$ in the above equations are non-universal constants 
depending on $g_c$ and the electron density of states. 
We see therefore that while the lowest dimension boundary perturbation leads to
the same 'singular' low temperature thermodynamic properties as for the 
underscreened  Kondo problem in a FL, the temperature dependence of the 
subleading terms which come from the electron tunneling term between the two 
channels with spin-flip scattering, are governed by the anomalous 
LL exponents and reflect the non-Fermi liquid nature of the system. 

If we add a magnetic field $H$, the residual spin impurity fluctuations are
suppressed  at low temperature $T<<H, H<< T_K$. Therefore the leading boundary perturbation is now the same as
that in the fully screened case, namely the electron tunneling operator between
the two channels with no spin-flip scattering (the term (iii) in the list of boundary operators) with dimension $(1+g_c)/2g_c$. The leading behavior of the 
thermodynamical quantities like the specific heat will be the same as in the 
fully screened Kondo case.\cite{furusaki,durga} In particular, the low 
temperature specific heat has a leading anomalous LL power law behavior 
instead of the logarithmic dependence while the impurity spin 
susceptibility shows to leading order the expected paramagnetic behavior.

\subsection{\label{sec:tCK} Zero backward Kondo scattering}
More generally, we can distinguish between forward and backward scattering strengths and express the interaction (Eq.\ref{eq:kondo}) as 
\bea 
 H_K = J_{K, f} [\psi _L^{\dag} (0) \frac{\vec \sigma}{2} \psi _L(0)  + \psi
 _R^{\dag} (0) \frac{\vec \sigma}{2} \psi _R(0) ]\cdot S \nonumber\\
+  J_{K, b} [\psi _L^{\dag} (0) \frac{\vec \sigma}{2} \psi _R(0) + \psi _R^{\dag} (0) \frac{\vec \sigma}{2} \psi _L(0)]\cdot\vec S.
\eea
For any generic values of $J_{K,f}, J_{K,b}$, the couplings flow to the strong 
coupling FP described above. But for $J _{K, b}=0$, the theory reduces to a two 
channel Kondo problem (with the left and right moving electrons corresponding to the two channels) interacting with a spin $S$ magnetic impurity. The charge sector 
decouples from the theory and it is sufficient to consider only the spin sector
to study the stability of the weak coupling fixed point. As is well known, the
weak coupling fixed point is stable for ferromagnetic (FM) coupling ($J _{K, f} >0$) while it is unstable for antiferromagnetic (AFM) Kondo coupling($J _{K, f} <0$). The Kondo temperature has the usual exponential coupling dependence. We 
can distinguish between three different cases for the low temperature physics.
While for $S=1/2$, the low temperature physics corresponds to the two channel  
overscreened Kondo physics, for $S=1$, the conduction electrons form a singlet 
with the impurity spin leading to fully screened Kondo physics, for $S>1$, the 
conduction electrons form a singlet with part of the impurity spin and the 
rest is left over as a decoupled spin of size $S-1$. For $S=1$, the low 
temperature physics exhibits the usual regular Fermi liquid behavior while for
$S>1$, the low temperature physics is governed by the marginally irrelevant 
ferromagnetic coupling between the residual spin of size $S-1$ and the 
conduction electrons. The latter again leads in the low temperature limit to 
the 'singular' Fermi liquid behavior (Eqs.\ref{eq:sing1}, \ref{eq:sing2}, \ref{eq:sing3}) described earlier.

\section{\label{sec:QD} Quantum Dot realizations}

\begin{figure}
\epsfig{figure=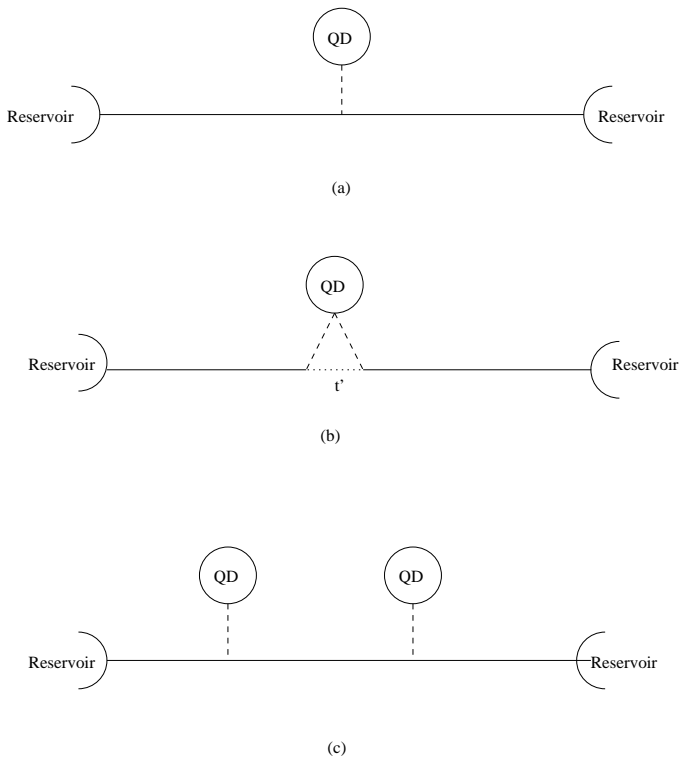,width=9cm,height=10cm}
\caption{\label{fig:geom}a) A quantum dot coupled to one site of a LL lead.
b) quantum dot coupled to two sites of a LL lead and c) Two 
quantum dots  attached to a LL lead}
\end{figure}

  We now discuss possible scenarios where we might observe such physics. One
  possible realization would be to couple a single quantum dot with spin $S$ to an interacting semiconducting wire (a LL wire)
 in the geometries shown in Fig.~\ref{fig:geom}a  and Fig.~\ref{fig:geom} b .  Another possibility is to couple two quantum dots with spin $S$ to a LL wire as shown in Fig.~\ref{fig:geom}c. 

\subsection{\label{sec:QD1}Quantum dot with spin S side-coupled to one site of a LL wire} 

 Let us first consider transport through the spin impurity realized as a QD 
coupled to LL lead in the geometry shown in Fig.~\ref{fig:geom}a. 
In the side coupled geometry, the Kondo effect appears as an anomalously 
strong reflection or backscattering rather than as transmission. At high 
temperatures, the weak coupling FP dictates the temperature dependence of the 
conductance and is essentially governed by the behavior of the backscattering 
Kondo scattering process. Therefore the high temperature linear conductance has the leading temperature dependence
 \be
 G(T) - G_0 \sim -G_0 S(S+1)(T/T_K) ^{(g_c-1)},
 \ee
where $G_0=2 e^2/h$ is the unitary conductance predicted in the absence of the
coupling to the QD. This is in contrast to the FL lead case where the conductance has the temperature dependence
 \be
 G(T) - G_0 \sim -G_0\frac{\pi ^2 S(S+1)}{4 (\ln (T/T_K))^2},
 \ee
due to the marginal nature of the Kondo exchange interaction. The 
leading temperature dependence of the conductance in the low temperature limit 
is governed by the hopping of an electron between the two semi-infinite LL 
leads via spin flip scattering and the electron tunneling operator with no spin
flip scattering. The low temperature conductance is therefore of the form:
 \be 
 G (T) \sim G_0 (a1 s(s+1) + a2)(T/T_K)^{\frac{1-g_c}{g_c}}.\label{eq:us_trans}
 \ee
where $a_1$ and $a_2$ are some non-universal constants.
This is in contrast to the FL lead case which shows a logarithmic 
temperature behavior
\be
 G (T) \sim G_0\frac{\pi ^2 s(s+1)}{4 (\ln (T_K/T))^2}.\label{eq:us_trans_FL}
\ee
Thus we find that the leading low temperature transport behavior is governed
by the subleading electron tunneling terms between the two channels without and with residual impurity spin-flip scattering and therefore shows non-Fermi 
liquid behavior with an anomalous power law behavior with the power law 
exponent being dictated by the LL interaction strength.

In a finite magnetic field, $T<<H<< T_K$, the leading temperature dependence
of the conductance is the same as that for the fully screened case. However,
since the dimension of the boundary operator governing the transport process in
 the two cases is the same, the conductance has the same temperature dependence
as for the underscreened case (see Eq.~(\ref{eq:us_trans})). The main difference between the two cases being the absence of the spin-dependent term.

\subsection{\label{sec:QD2}Quantum dot with spin S side-coupled to two sites of a LL wire} 

 We next consider electronic transport through a dot configuration where 
the dot is coupled to two different sites of the LL chain as shown in 
Fig.~\ref{fig:geom}b. When the electron density is at half-filling and for $1/2<g<1$, it is well known that the charge sector in the LL model becomes massive
 but the spin sector still remains massless. It can be then shown~\cite{furusaki,johan} that this model is a realization of the case $J_{K,b} =0$ discussed at 
the end of the previous section. As discussed in the previous section, the 
problem becomes then essentially that of a two channel Kondo problem with a 
spin $S$ impurity. Such a model has been previously studied~\cite{eggert}. 
For $S=1/2$, the low temperature transport exhibits overscreened Kondo 
behaviour:
\be
G(T) \sim G_0(T/T_K)^{1/2}. 
\ee
For $S=1$, the transport shows FL behavior
\be
G(T) \sim G_0 (T/T_K)^2,
\ee
while for $S>1$, the conductance shows the underscreened behavior given in
Eq.~(\ref{eq:us_trans_FL}) with $s=S-1$.

For electron densities away from half filling, the problem can be thought of 
that of a quantum dot in an embedded geometry \cite{simon3} with potential 
scattering. A similar analysis as in the $S=1/2$ case ~\cite{fabrizio, simon3} 
shows that for $0<g_c<1/2$, the strong coupling fixed point is the same as that
for the two channel Kondo FP with a spin $S$ impurity, while for $1/2<g_c<1$, 
one obtains the strong coupling FP of the single channel LL with a spin $S$ impurity. We note that this implies that in contrast to non-interacting electrons, 
one will not get a Kondo resonance in general for interacting electrons (for $1/2<g_c<1)$ except at some particular value of the gate voltage where the 
backscattering term vanishes. Off resonance, the low $T$ conductance has the 
same behaviour as in \ref{eq:us_trans}. For gate voltages very close to the
resonance voltage, $G(T) -G_0 \propto -G_0 (a_3 + a-4(s(s+1) T^{(1-g_c)}$ where $a_3$ and $a_4$ are non-universal constants. We note that such a scaling 
behaviour was observed in earlier studies of the gate voltage dependence of the
linear conductance through a Kondo spin $1/2$ quantum dot coupled to LL leads.~\cite{meden}

\subsection{\label{sec:QD3}Two spin S quantum dots side-coupled to a LL wire} 

Another possibility is to couple two quantum dots with spin $S$ to a LL wire as
shown in Fig. \ref{fig:geom}c. The latter problem is equivalent to that of 
two magnetic impurities in a LL.\cite{egger1,egger2, fulde} When there are 
more than 
one magnetic impurity, there are two competing effects: the Kondo spin exchange
 interaction between each impurity spin and the conduction electron spin and 
the induced Ruderman-Kittel-Kasuya-Yosida (RKKY) spin exchange interaction between the impurities (the RKKY 
interaction is modified in the presence of electron interaction \cite{egger1}).
  The ground state of the system depends on which of these interactions dominate.If 
the Kondo interaction
strength is greater than the RKKY interaction, then one expects single 
impurity physics. The impurity spin then forms a singlet with the conduction 
electrons and gets decoupled. The electrons then see only an effective potential scatterer at each impurity site. So effectively, for two impurities, the chain behaves like as if there are two barriers. Generically, one should expect the
zero temperature conductance to be zero. However, there is the interesting 
possibility of resonant tunneling in the Kondo limit (for not very large distances between the two impurities and if the resonant tunneling conditions are 
satisfied) just like for symmetric double barriers.\cite{kane}  On the other 
hand, if the 
RKKY  interaction dominates, there can be different kinds of physics depending 
on whether there is FM or AFM interaction between the two impurities. For AFM
coupling between the spins, one expects the two impurities to lock into an 
effective singlet state which is essentially like a non-magnetic impurity. In 
the side coupled configuration, one expects the non-magnetic impurity to have 
no effect on the conduction electrons and therefore lead to the unitary value
for the zero temperature conductance. Thus, while the Kondo limit and AFM 
exchange limit both show a singlet phase, they exhibit different physics in that
in the Kondo limit, one expects `breaking' of the chain except under some circumstances where resonant tunneling can occur while in the RKKY AFM limit, one
expects 'healing' of the chain. For strong FM RKKY interaction, the problem 
effectively becomes that of a spin $2S$ impurity interacting with a LL, the 
problem therefore becomes effectively the underscreened Kondo problem discussed in the previous sections. The low temperature conductance then has the 
temperature dependence given in Eq.~(\ref{eq:us_trans}) reflecting LL behavior.
We also mention that recent experiments~\cite{RKKYexpt1, RKKYexpt2} on quantum 
dots with a non-local RKKY interaction have motivated studies of transport in 
such coupled quantum dot systems.\cite{tamura,simon1,simon2,durga2} However, 
these studies do not consider the effect of electron-electron interactions.

\section{Conclusion} 
To summarize, we have analyzed the problem of underscreened Kondo physics in a LL.
 We find that the leading temperature dependence of thermodynamical quantities 
like the specific heat, spin susceptibility are FL like in nature. 
However, the anomalous LL power law exponents are seen in the subleading terms. 
We have also discussed possible 
realizations through single and double quantum dot configurations coupled to LL
leads and the consequences for electronic transport. The leading low 
temperature transport behavior is seen to exhibit in general, non Fermi 
liquid LL behavior unlike the thermodynamical quantities. 

\begin{acknowledgments}
P.D would like to thank DST, India for financial support. P.D also
acknowledges partial support from IFCPAR-CEFIPRA(Project No.3104-2), New Delhi
 and thanks CEA, Grenoble where part of this work was done. PS acknowledges interesting discussions with I. Affleck and thanks PITP-Vancouver where part of this work was done.  

\end{acknowledgments}

\end{document}